\documentclass[a4paper,11pt]{article}
\pdfoutput=1 

\usepackage{jinstpub} 

\title{Multi-contrast K-edge imaging on a bench-top photon-counting CT system: Acquisition parameter study}

\author[a,1]{Devon Richtsmeier}
\author[a]{Chelsea A. S. Dunning}
\author[b]{Kris Iniewski}
\author[a]{Magdalena Bazalova-Carter}

\affiliation[a]{Department of Physics and Astronomy, University of Victoria,\\ Victoria, V8P 5C2, Canada}
\affiliation[b]{Redlen Technologies,\\ Saanichton, V8M 1X6, Canada}

\emailAdd{drichts@uvic.ca}

\abstract{

\emph{Purpose:} Photon-counting computed tomography (PCCT) shows promise for medical imaging in regards to material separation and imaging of multiple contrast agents. However, many PCCT setups are under development and are not optimized for specific contrast agents or use cases. Here, we demonstrate how experimental system parameters may be varied in order to enhance performance and we propose a set of recommendations to achieve this based on contrast agent.

\emph{Approach:} A table-top PCCT system with a cadmium zinc telluride (CZT) detector capable of separating six energy bins was used to image multiple contrast agents in a small phantom. The contrast agents were separated and the concentration was quantified using K-edge subtraction. To increase system performance, we investigated three parameters: beam filter type and thickness, projection acquisition time, and energy bin width. The results from the parameters were compared based on PCCT signal and contrast to noise ratio (CNR) or noise in K-edge images. The concentrations of the contrast agents were quantified in K-edge images and compared to known concentrations.

\emph{Results:} The bench-top PCCT system was able to successfully quantify the contrast agents through K-edge subtraction. Decreasing projection acquisition time showed a decrease in K-edge CNR. However, it did not scale as the square root of time. Filter type and bin width demonstrated a dependence on the specific contrast agent.

\emph{Conclusions:} The presented bench-top system demonstrated the ability to separate contrast agents using K-edge subtraction and accurately determine contrast concentration in K-edge images. Specific parameters for future use will be chosen based on contrast agent.
}


\begin{document}
\maketitle
\flushbottom

\section{Introduction}
\label{sect:intro}

Computed tomography (CT), first described by Hounsfield and others \cite{Hounsfield1973, Cormack1963, Cormack1964}, has been widely used for research \cite{Garcea2018, Paulus2000, Pereira2020} and in clinical imaging \cite{Nieman2001} since the first CT scanner was built \cite{Hounsfield1980}. While conventional CT offers many benefits such as high spatial resolution and fast imaging time, one drawback is its inability to distinguish between materials with similar attenuation curves, such as those with effective atomic numbers (Z) that are close to one another. Energy selective CT, or multi-energy CT, was first proposed in the 1970's \cite{Alvarez1976}, allows for material decomposition while still offering the benefits of conventional CT. Energy selective CT functions by utilizing multiple beams with different energies, or a single beam with detectors able to separate a single beam into two or more energy spectra to obtain energy information from the detected x-rays. The most widely used type of energy selective CT is dual energy CT (DECT), in which the object being imaged is typically scanned with two beams of different energies. The two beams result in different images, which can be exploited to separate similar materials. Clinically, DECT is used to distinguish materials based on their energy attenuation dependencies. Using DECT, materials such as bone and iodine (a common CT contrast agent) can be separated \cite{Chong2008} or the amount of injected contrast can be quantified. For example, in order to determine the composition of kidney stones \cite{Graser2009}, DECT can be used to quantify the concentration of iodine present in the image. A number of algorithms and methods have been developed to accomplish such material decomposition and quantification. Bazalova \textit{et al} demonstrate how materials can be separated by converting the measured Hounsfield units (HU) into effective atomic number (Z) and and the corresponding electron density by iteratively solving for Z \cite{Bazalova2008}. Mendonca \textit{et al} provide an algorithm and show how multiple materials can be distinguished through the identification of multiple linear attenuation coefficient triplets \cite{Mendonca2014}. 
		
With the advent of high-flux photon-counting detectors (PCDs) \cite{Fischer2000, Lindner2001, Iwanczyk2009}, research into the uses of multi-energy CT has been underway. Conventional CT and DECT utilize energy integrating detectors (EIDs), which detect x-rays by measuring the total energy deposited in the detector without accounting for the individual energy deposited by each x-ray. PCDs allow for the spectral binning of x-rays into predefined energy ranges by measuring the pulse height of every interaction in the semiconducting layer of the detector \cite{Willemink2018}. Imaging modalities like positron-emission tomography (PET) rely on energy discriminating detectors to detect 511 keV photons with high resolution, but those detectors operate at much lower flux rates than the beams produced by CT scanners \cite{Shukla2006}. With the development of high-flux PCDs, photon-counting CT (PCCT) is possible, in which only one beam energy is necessary. A number of groups are even working towards full-body clinical PCCT scanning \cite{Ronaldson2012, Muenzel2017, Persson2014}. 
    	
The PCDs used for PCCT normally have a number of energy thresholds which determine the edges of the energy ranges, or bins, that individual x-rays are placed into. These energy thresholds are tunable, which allows for the specific energy ranges to be isolated or for the energy threshold to be placed at specific energies corresponding to the K-edge of a contrast material. The K-edge of an element is characterized by a large increase in the attenuation of photons with energies above the K-edge. This is because x-rays with energies above the K-edge are able to eject electrons in the K-shell of the atoms of that element. Placement of a threshold at a K-edge gives the ability to take advantage of this increase in attenuation to distinguish high-Z contrast materials from the background using K-edge subtraction. K-edge subtraction is accomplished by the acquisition of a higher energy image above the K-edge and a lower energy image below the K-edge and subtracting the lower from the higher. A detailed description of the algorithm can be found elsewhere \cite{Zhang2020}. In the two images, the signal from the background will be effectively constant because the attenuation properties of the background vary only slightly over the energy range around the K-edge. However, the difference in attenuation of the contrast material above and below the K-edge is large due to the increase in photoelectric absorption. This results in leftover signal only in areas containing the contrast material. K-edge subtraction has been used with mono-energetic x-rays \cite{Sarnelli2004, Thomlinson2018} to acquire the two images above and below the K-edge as well as with PCCT systems \cite{Shikhaliev2012, Zhang2020}.
        
PCCT is actively being researched for use in areas of preclinical imaging research such as cancer treatment \cite{Moghiseh2018} and the imaging and separation of contrast agents \cite{Si-Mohamed2017, Cormode2017, Symons2017, Dunning2020, Wang2011a} as well for use clinically \cite{Ding2014, Cho2015, Yu2016, Willemink2018}. In both the clinic and other areas it is necessary to develop protocols which offer the best results based on the work or research being performed. For example, in small animal research, the imaging procedure needs to be optimized in order to have an imaging time that reduces movement artifacts and stress to the animals as well as keeping radiation dose to the animals at a minimum. However, reducing imaging time too drastically can have detrimental effects on image quality. In the clinic, dose and imaging time also need to be kept low in order to reduce patient motion, discomfort, and future risk of cancer while also keeping image quality high enough for accurate diagnosis or treatment planning. 
        
One current area of research using PCCT is imaging of multiple contrast agents \cite{Symons2017, Cormode2017, Wang2011a}. Depending on the number of energy thresholds available, PCCT offers the ability to accomplish this using K-edge subtraction. The potential benefit this offers is to distinguish multiple tissues or processes in one scan, thereby reducing dose and allowing for the possibility of studying how different processes may interact \textit{in vivo}. However, the PCCT system used must be optimized for the best results depending on the desired image quality and contrast agent(s) used. In this work we demonstrate the experimental variation of a few parameters on a bench-top PCCT system to obtain the best results as measured by the resulting signal and K-edge contrast-to-noise ratio (CNR). The adjusted parameters were filter type and thickness, projection acquisition time, and energy bin width due to their effect on image quality in previous research \cite{Roessl2006, Ren2016, Wang2009}. A set of practices was developed to choose the correct parameters in our system based on the contrast agent.

\section{Materials and methods}

\subsection{Phantoms and imaging setup}

\begin{figure}
\begin{center}
\begin{tabular}{c}
\includegraphics[width=14cm]{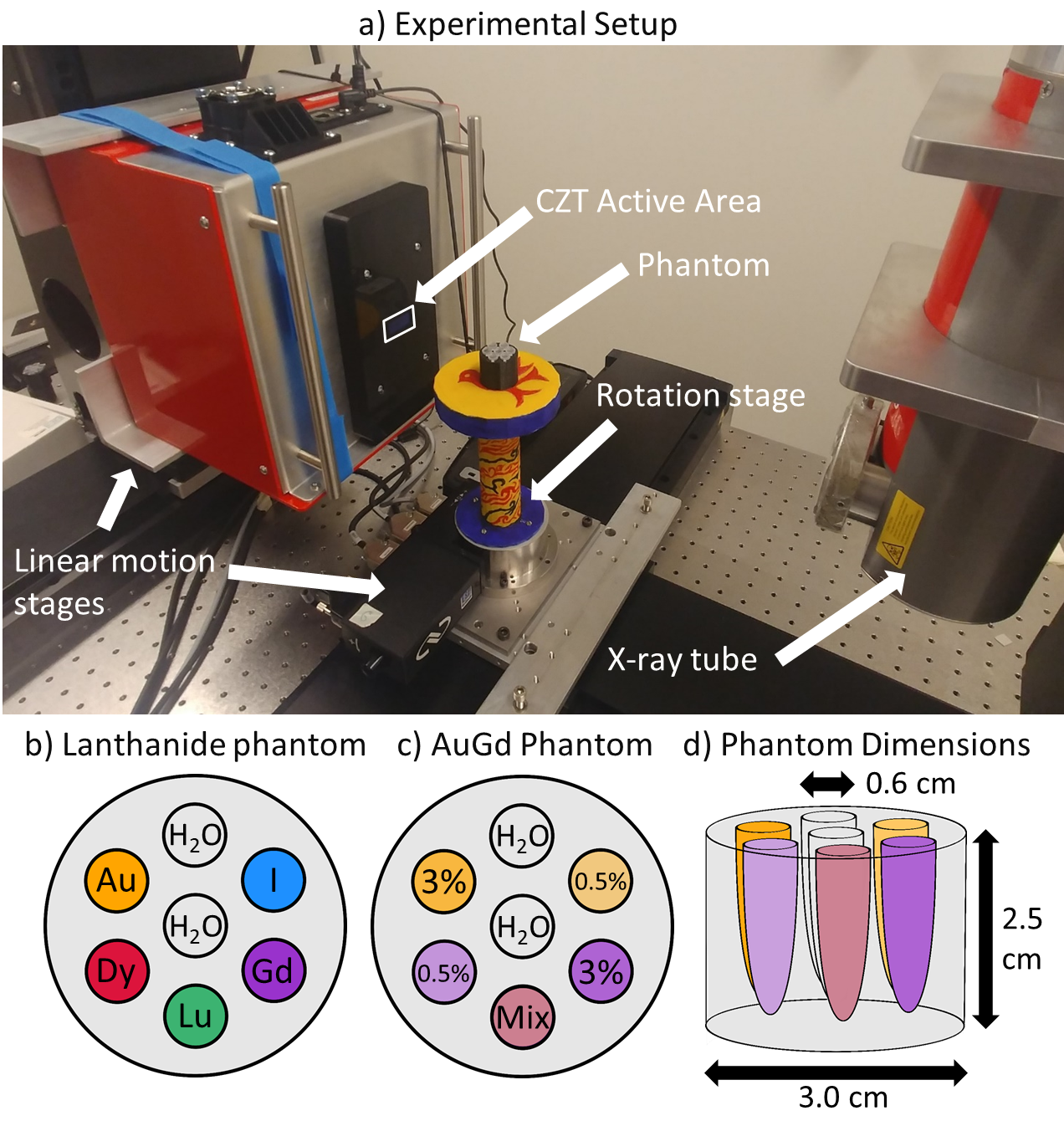}
\end{tabular}
\end{center}
\caption 
{ \label{fig:setup}
\emph{Experimental setup.} a) Experimental setup showing the x-ray tube, detector, phantom, and movement stages. b) Contrast material layout for the three Lanthanide phantoms, one for each concentration of 5\%, 3\%, and 1\% of all contrast materials. c) Layout for the AuGd phantom showing \% concentration in each vial with Au (orange) and Gd (violet). Mix contained 0.5\% concentration of both materials. d) The phantom dimensions.} 
\end{figure}

The imaging setup and phantoms used in this study are shown in figure~\ref{fig:setup}. The imaging setup consisted of an x-ray tube, rotation stage, various linear motion stages, the phantom containing the various contrast agents, and a cadmium zinc telluride (CZT) detector. The x-ray tube was an XRS-160 (Comet Technologies, San Jose, CA) mounted on two linear motion stages (Newport Corporation, Irvine, CA) allowing it move vertically as well as away from and towards the detector. The CZT detector module (Redlen Technologies, Saanichton, BC, Canada) consisted of the 8$\times$24 mm$^2$ sensor (shown as the Active Area in figure~\ref{fig:setup}a) and the housing with the hardware and cooling system. The sensor is 2 mm thick with a 330 $\mu$m pixel pitch and is capable of operating at 250 Mcps/mm$^2$ without polarization \cite{Iniewski2016}. The high-flux detector operates in non-paralyzable mode and the pile-up effects are smaller than 2\% with the parameters used here \cite{Iniewski2019}. The detector also has 6 energy thresholds that can be tuned by the user. To calibrate the thresholds, the central energy threshold was swept over the entire energy range twice to obtain the spectra of Co-57 and Am-241, both of which have a distinct energy peak. Each pixel was then calibrated based on their response. The CZT detector was mounted on two linear motion stages letting it move both vertically and parallel to the plane of the detector. The rotation stage (Newport Corporation, Irvine, CA) was itself mounted on a motion stage between the tube and the detector which allowed for motion parallel to the plane of the detector as well. 

\begin{figure}
\begin{center}
\begin{tabular}{c}
\includegraphics[width=10cm]{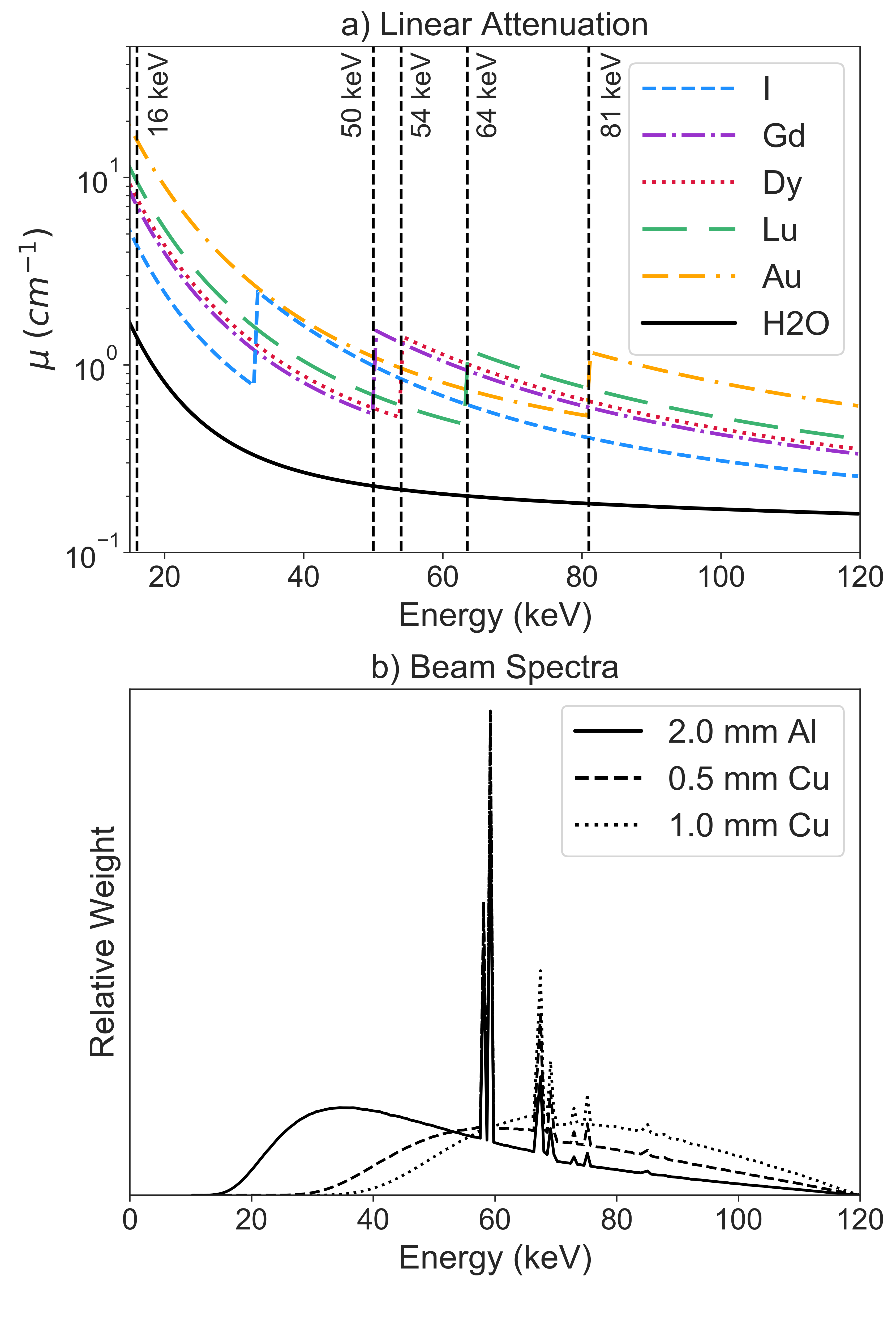}
\end{tabular}
\end{center}
\caption 
{ \label{fig:Attenuation}
\emph{Linear attenuation coefficients and beam spectra.} a) Semi-log plot of the linear attenuation coefficients of water and all contrast agents (5\% concentrations) included in the study, over the relevant energy range. b) Relative number of photons with respect to energy of the three filtered beam spectra. Each spectra was scaled by the tube current used for that filtration.} 
\end{figure}

The phantom (figure~\ref{fig:setup}d), was 3D printed from solid polylactic acid (PLA) with a density of 1.25 g/cm$^3$ (3D Hubs, Victoria, BC, Canada). The phantom has seven holes with a diameter of 0.6 cm which fit Fisherbrand 0.2 ml PCR tubes (Fisher Scientific, Nepean, ON, Canada). These tubes (or vials) contained pure water or water-based solutions of gold, gadolinium, dysprosium, lutetium, or iodine of various concentrations. Four phantom configurations were used in the study, three of which followed the layout shown in figure~\ref{fig:setup}b, (called the Lanthanide phantoms), with concentrations of 5\%, 3\%, or 1\% for all materials in that particular phantom. The fourth phantom, hereby refered to as the AuGd Phantom, is shown in figure~\ref{fig:setup}c, with gold and gadolinium concentrations and with the mixture vial consisting of 0.5\% concentrations of both materials. The concentrations were chosen in order to be sure they were visible with the current bench-top system. The gold solutions were synthesized from gold (III) chloride (GG3CS-25.4-100 Lot AUY03-7077, Nanopartz Inc, Loveland, CO). The iodine solutions used the iodine-based Omnipaque 300 (iohexol, GE Health-care, Princeton, NJ). Gadolinium, dysprosium, and lutetium chloride hexahydrate salts (Sigma-Aldrich, Oakville, ON) were used to synthesize the remaining solutions by dissolving the salts of the specific element in water to obtain a 5\% concentration by weight. This stock was then diluted as necessary. The purities of the salts were 99.99\% for LuCl$_3$, 99.9\% for DyCl$_3$, and 99.999\% for GdCl$_3$. The plots of each contrast agent's linear attenuation coefficient (at 5\% concentration) with respect to x-ray energy can be seen in figure~\ref{fig:Attenuation}a. 

\subsection{Data acquisition}

Filter type and thickness, projection acquisition time, and bin width were examined with different phantoms and setup parameters. All images were acquired with a cone-beam geometry with the phantom at isocentre, located 32 cm from the x-ray source and 11 cm from the detector \cite{Dunning2020a}. The CZT detector was connected to a high-speed photon counting Application Specific Integrated Circuit (ASIC), operating at rates up to 62.5 Mcps per channel. The ASIC allowed photons incident on the detector to be sorted into six energy bins with the thresholds set by the user and varied in this study between the different acquisitions. The sixth energy bin was used for overflow counts of the detector in all scans.

Each phantom was scanned by acquiring 180 projection images in 2 degree intervals over a 360$^{\circ}$ rotation about the isocentre. Three separate 360$^{\circ}$ rotations were performed for each phantom in order to image the entire phantom body. Between each rotation, the detector was translated 13.5 mm parallel to the detector plane to image the full phantom with the small-area detector. The projections from each rotation were then combined to produce the entire projection data set of the full phantom. 

All acquisitions were captured with a tube voltage of 120 kVp using the small focal spot (1.0 mm). The tube current varied between different setups, and is detailed in each of the evaluated parameter subsections below. Unless specified otherwise, one projection image was captured over 1 second every two degrees, and the phantom was rotated at a speed of 1.48 $^{\circ}$/s to accommodate for the time to transmit the data from each projection. This resulted in an imaging time of approximately 4 minutes per rotation and $\sim$12 minutes total.

\subsubsection{Filter selection}

Three different filter setups were investigated for this study: 2.0 mm Al, 0.5 mm Cu, and 1.0 mm Cu. In order to obtain similar noise levels in the final images using each filter, the tube current was varied to acquire a similar total number of photons in the airscan when compared with the total number when using a 1.0 mA tube current with 2.0 mm Al filtration. This yielded tube currents of 2.25 mA and 4.75 mA for 0.5 mm Cu and 1.0 mm Cu, respectively. The spectra for each beam filtration can be seen in figure~\ref{fig:Attenuation}b. The unfiltered spectrum was calculated using a validated EGSnrc Monte Carlo model and then filtered according to Beer’s Law. The three filters were chosen due to the use of aluminum and copper in other CT and PCCT studies \cite{Ren2016, Persson2014, Dunning2020, Ronaldson2012, Roessl2006} and to cover a lower, middle, and higher weighted energy spectrum.

All three Lanthanide phantoms were scanned using the tube currents listed above. Bin thresholds were set to 16, 50, 54, 64, 81, and 120 keV to accommodate the K-edges of gadolinium (50.2 keV), dysprosium (53.8 keV), lutetium (63.3 keV), and gold (80.7 keV). This translated to bins with energy ranges of 16--50 keV, 50--54 keV, 54--64 keV, 64--81 keV, and 81--120 keV. 

\subsubsection{Bin width selection}

In order to facilitate the modification of bin width, only two contrast agents were investigated, placed in the AuGd phantom. 0.5 mm Cu filtration and a tube current of 2.25 mA were used for image acquisitions as it offered a compromise between the other two filters detailed above. The energy thresholds were shifted to create different bin widths on either side of the K-edge of the two contrast materials. The two energy thresholds that remained the same were those set at 50 keV and 81 keV, corresponding to the K-edges of gadolinium and gold. Four different scans were taken of the AuGd phantom corresponding to bin widths of 5, 8, 10, and 14 keV for gadolinium and bin widths of 5, 10, 14, and 20 keV for gold. 

\subsubsection{Projection time selection}

Three different times were investigated: 1 s, 0.5 s, and 0.1 s. While 1 s acquisitions gives adequate image quality, lower acquisition times were explored to examine image quality at lower imaging doses. For all three options, 0.5 mm Cu filtration was used along with a tube current of 2.25 mA. The 3\% Lanthanide phantom was used, with identical energy thresholds as those used for filter selection, except that the lowest threshold was moved from 16 keV to 35 keV to exclude the lower energy photons which are most affected by charge sharing.

For the projection acquisition, 1800 0.1 s projections were collected while the phantom was rotated 360$^{\circ}$ at a speed of 0.471$^{\circ}$/s. 180 projections were used for image reconstruction with every ten projections summed together for 1 s acquisition, every other five acquisitions summed for 0.5 s acquisitions, and every tenth acquisition was used for 0.1 s acquisitions. The imaging time per rotation was approximately 12.5 minutes for a total imaging time of 37.5 minutes.

\subsection{Dose}
The dose to the phantom was calculated at each filter and tube current setup with 1 s projection acquisitions by simulating a cylindrical water phantom in TOPAS \cite{Perl2012} as described by Dunning \textit{et al} \cite{Dunning2020a}. For both 2 mm Al and 1 mm Cu filtration, with their corresponding tube currents, the dose to the phantom was 333 mGy. For 0.5 mm Cu and 2.25 mA tube current, the dose was 250 mGy.

\subsection{Image reconstruction and analysis}

\subsubsection{PCCT image reconstruction}

All images were reconstructed using MATLAB (The Mathworks, Natick, MA) and analyzed using Python. Projections were created by converting the count data from the detector using the following equation:

\begin{equation}
\label{eq:intensity}
p_n = -ln\Big(\frac{I_n}{I_{0,n}}\Big);\;\;  n=1, 2, 3, ..., 7\, 
\end{equation}

where $I$ is the intensity of the beam including the phantom and $I_0$ is the intensity of the beam in air. $I_0$ was obtained from an airscan taken at each corresponding detector translation in order to normalize the count data ($I$) to the flat field. The index number $n$ corresponds to the index number of each of the six energy bins and the full count data bin ($n=7$). The CT images from each of the corrected projection data sets were reconstructed using the Feldkamp-Davis-Kress algorithm \cite{Feldkamp1984} with a Hamming filter.

\subsubsection{K-edge images}

K-edge images were created by subtracting the CT image of the energy bin below each respective K-edge from the CT image of the energy bin above the K-edge. For example, to create the K-edge image of gadolinium using the AuGd phantom and 5 keV bin widths, the image of the 45--50 keV bin would be subtracted from the 50--55 keV bin.

\subsubsection{Image analysis}

Analysis of the CT images was done by first normalizing all of the voxels to Hounsfield units (HU) by applying the following equation to each voxel:

\begin{equation}
\label{eq:HU}
HU = 1000*\frac{\mu - \mu_{water}}{\mu_{water}}\, ,
\end{equation}

where $\mu$ is the signal in each voxel and $\mu_{water}$ is the average signal in the water vial. The water vial signal was obtained separately for each filter choice and used for all subsequent images taken with the same parameters. To obtain signal for each of the contrast agents at the various concentrations in each of the phantoms, a region of interest (ROI) encompassing only the inside of each vial was drawn and the mean value of the voxels within the ROI was taken to get the average signal ($\mu_{ROI}$) and the standard deviation of the voxel values ($\sigma_{ROI}$) was calculated to evaluate the error bars.

Analysis of the K-edge images was done in two ways. First was to calculate the CNR of each contrast agent in its respective K-edge image and the error (CNR$_{err}$) using the following equations:

\begin{equation}
\label{eq:CNR}
\begin{split}
CNR = \frac{\mu_{ROI} - \mu_{phantom}}{\sigma_{phantom}} \,,
\qquad
CNR_{err} = \frac{\sqrt{\sigma^2_{ROI} + \sigma^2_{phantom}}}{\sigma_{phantom}}\,,
\end{split}
\end{equation}

where $\mu_{phantom}$ and $\sigma_{phantom}$ are the mean and standard deviation of the phantom body, respectively.

Second, to calculate the concentration associated with the measured K-edge signal, the average signal from the ROI of the contrast material at 5\% was measured in one set of images from one acquisition at each filter choice and used for image normalization in all phantoms for subsequent acquisitions with the same parameters. Then, the average signal was taken in an ROI to find the measured concentration.

\section{Results}

\subsection{Filter selection}

\begin{figure}
\begin{center}
\begin{tabular}{c}
\includegraphics[width=12cm]{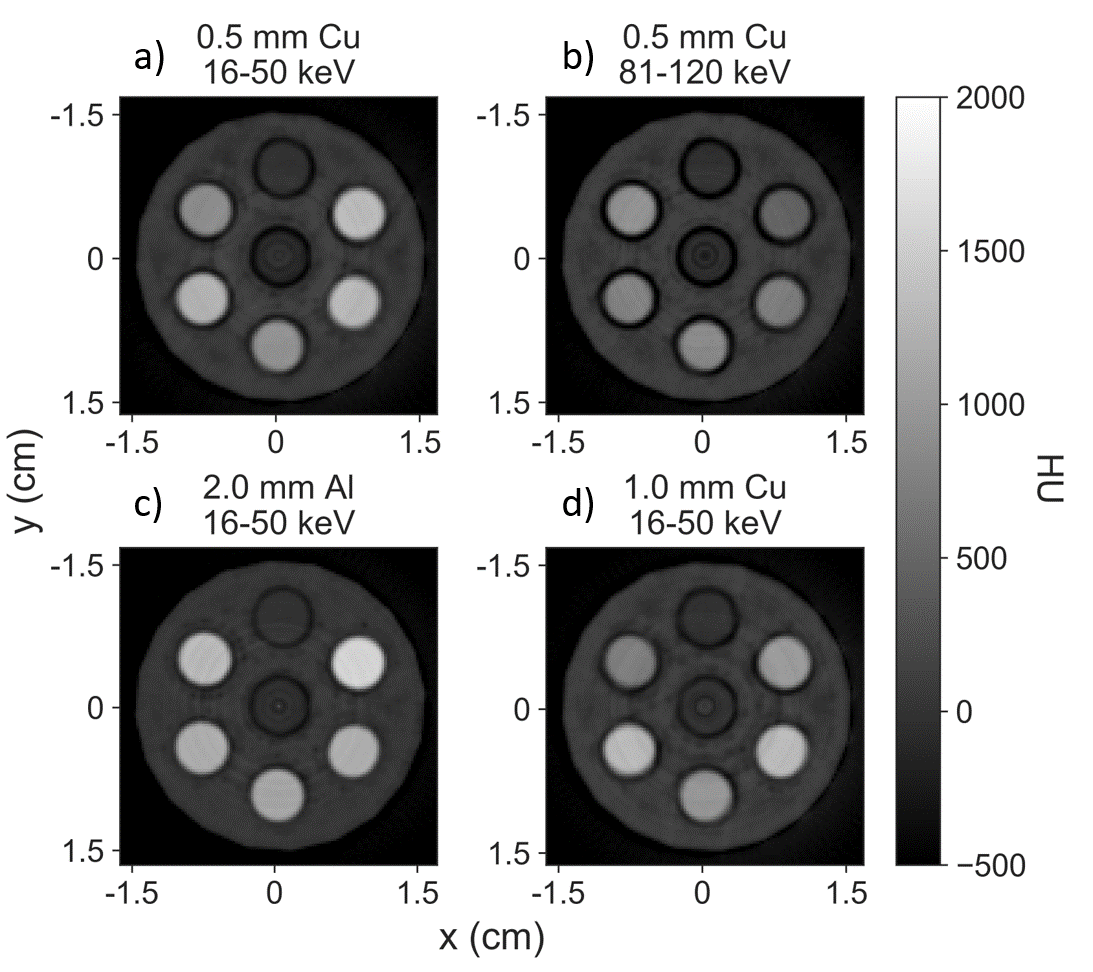}
\end{tabular}
\end{center}
\caption 
{ \label{fig:Filter CT}
\emph{CT images from filter selection}. a) A CT slice showing the 3\% lanthanide phantom with 0.5 mm Cu filtration in the 16--50 keV energy bin. b) The same CT slice of the same phantom with 0.5 mm Cu filtration in the 81--120 keV energy bin. c) A similar slice of the same phantom with 2.0 mm Al filtration in the 16--50 keV energy bin. d) A similar slice with 1.0 mm Cu filtration in the 16--50 keV energy bin.} 
\end{figure}

\begin{figure}
\begin{center}
\begin{tabular}{c}
\includegraphics[width=14cm]{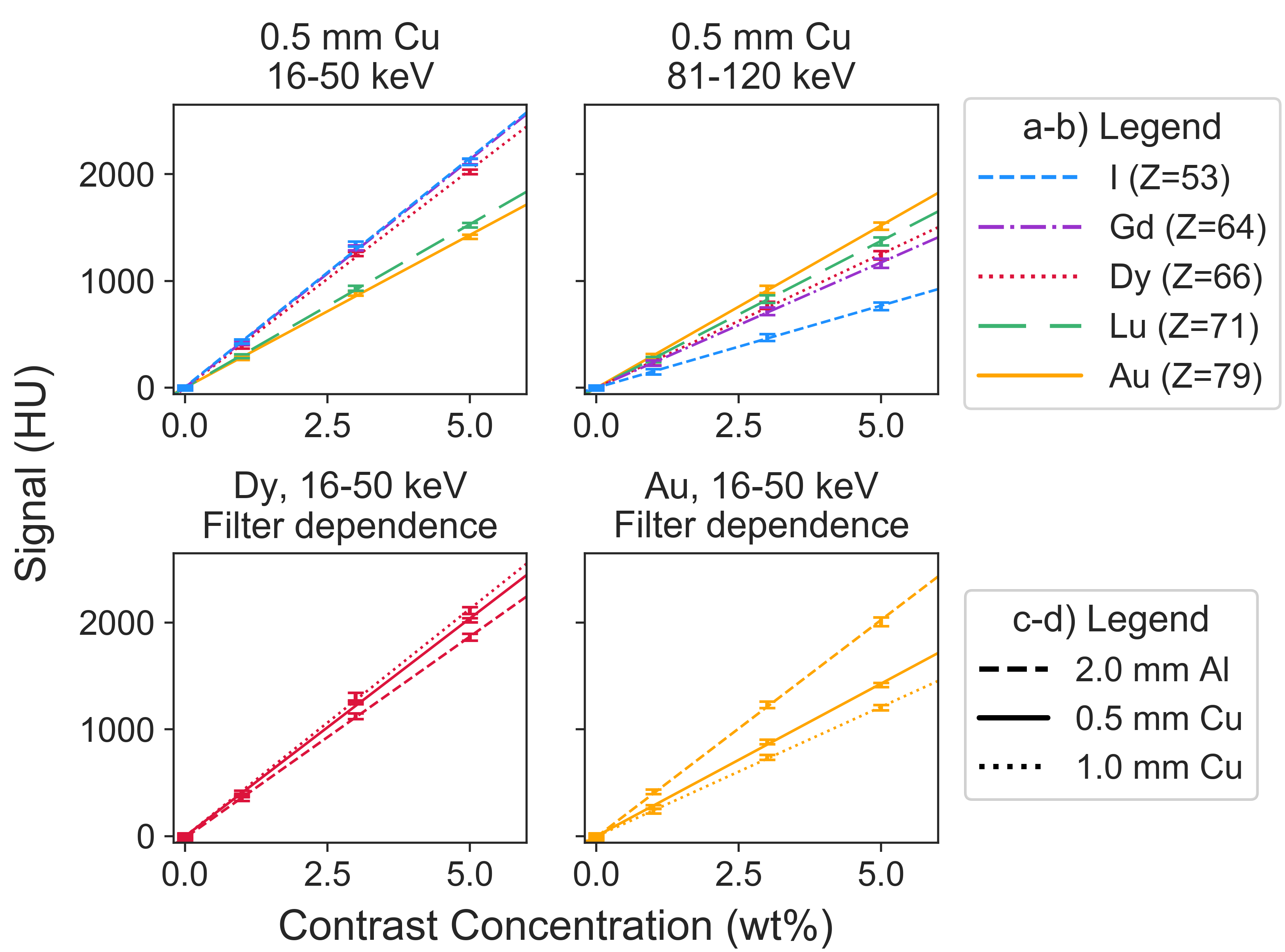}
\end{tabular}
\end{center}
\caption 
{ \label{fig:Filter}
\emph{Filter selection.} a) Signal from all contrast agents at 1\%, 3\%, and 5\% concentrations demonstrating a linear relationship. Energy range of 16--50 keV with 0.5 mm Cu filtration. b) The same components as in a), instead showing the signal in the energy range 81--120 keV. c) Gadolinium signal at all 3 concentrations demonstrating the effect of three different filters in the 16--50 keV energy range. d) The same setup as in c) with gold replacing gadolinium as the contrast agent.} 
\end{figure}

Examples of the CT images analyzed for this study are shown in figure~\ref{fig:Filter CT}. Figure~\ref{fig:Filter}a--d show the results of the filter selection for various concentrations of each of the contrast agents. The corresponding CT images from which the data was analyzed can be seen in figure~\ref{fig:Filter CT}a--d. Figure~\ref{fig:Filter}a and b show CT signal as HU vs. contrast concentration for all contrast elements with 0.5 mm Cu filtration. A linear signal response with contrast concentration is seen, with the slope of the line decreasing with increasing Z for the lowest energy bin (figure~\ref{fig:Filter}a) and the slope increasing with increasing Z for the highest energy bin (figure~\ref{fig:Filter}b). Figure~\ref{fig:Filter}c shows the linear relationship of signal with concentration for dysprosium with each filter in the 16--50 keV bin, in which the slope of the lines decreases from 1.0 mm Cu, to 0.5 mm Cu, and finally to 2.0 mm Al. Figure~\ref{fig:Filter}d demonstrates the same lines as figure~\ref{fig:Filter}c but for gold in the 16--50 keV bin, with the slopes of the lines reversing their order from those shown in figure~\ref{fig:Filter}c.

\begin{figure}
\begin{center}
\begin{tabular}{c}
\includegraphics[width=14cm]{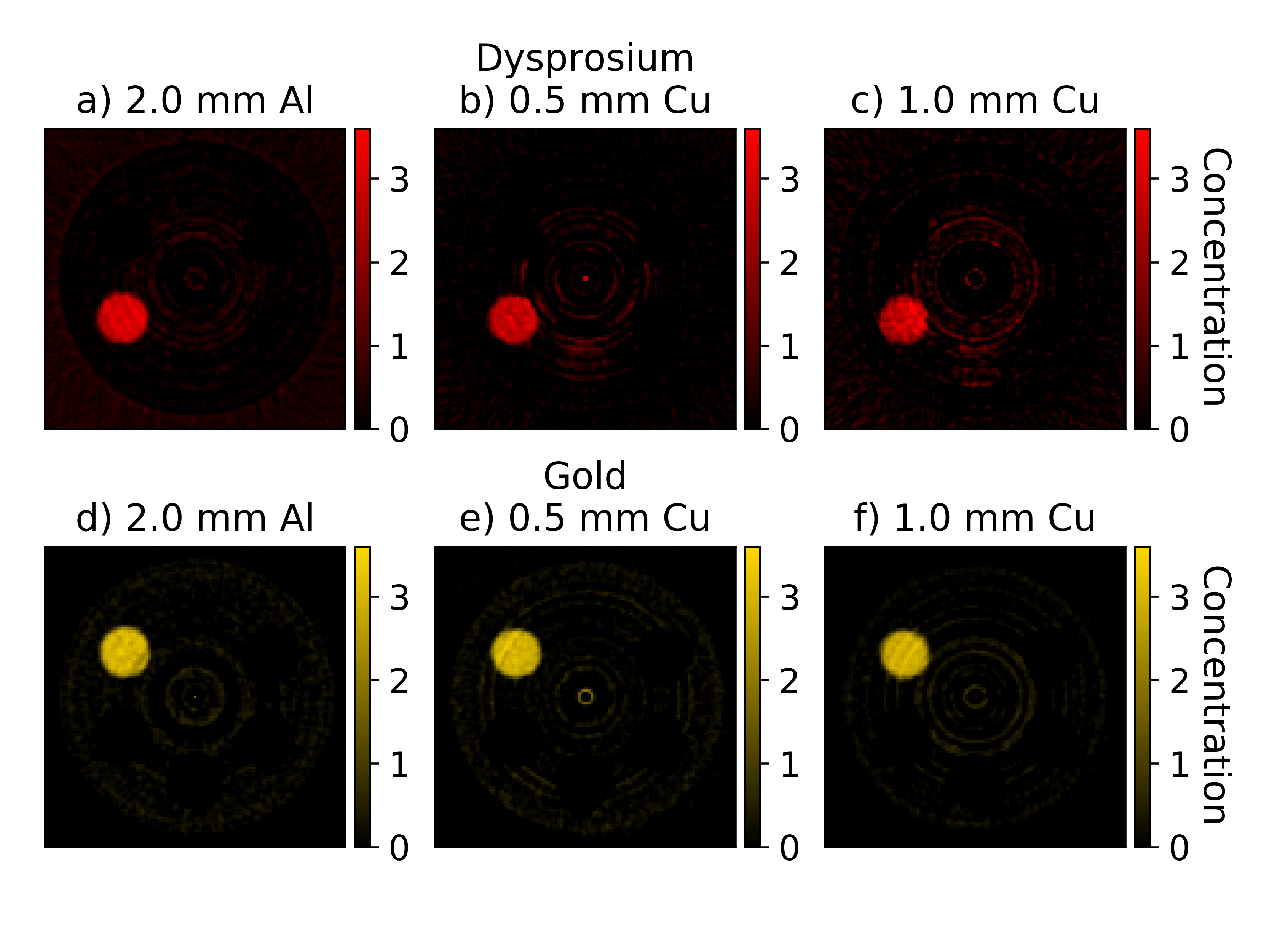}
\end{tabular}
\end{center}
\caption 
{ \label{fig:Filter K-Edge}
\emph{K-edge images with different filters.} a--c) K-edge images of 3\% dysprosium with the different filters: a) 2.0 mm Al, b) 0.5 mm Cu, and c) 1.0 mm Cu. d--f) K-edge images of 3\% gold with the different filters: d) 2.0 mm Al, e) 0.5 mm Cu, and f) 1.0 mm Cu.} 
\end{figure} 

Figure~\ref{fig:Filter K-Edge} depicts K-edge images using the three filters for dysprosium (a--c) and gold (d--f), both at 3\% concentration. For dysprosium, an increase in noise is seen for increasing filtration from 2.0 mm Al to 0.5 mm Cu and finally to 1.0 mm Cu. For gold the opposite is true, with noise decreasing with softer filtration. Table~\ref{tab:filterCNR} shows the K-edge CNR for all four contrast agents with each filter type.

\begin{table}[ht]
\caption{K-edge CNR by contrast agent (3\%) and filter type.} 
\label{tab:filterCNR}
\begin{center}       
\begin{tabular}{|c|c|c|c|c|} 
\hline
\rule[-1ex]{0pt}{3.5ex}  Filter & \multicolumn{4}{c|}{Contrast agent} \\
\cline{2-5}
\rule[-1ex]{0pt}{3.5ex} type & Gd ($Z=64$) & Dy ($Z=66$) & Lu ($Z=71$) & Au ($Z=79$) \\
\hline
\rule[-1ex]{0pt}{3.5ex}  2.0 mm Al & 3.5 $\pm$ 1.4 & 10.6 $\pm$ 1.3 & 19.9 $\pm$ 1.3 & 12.6 $\pm$ 1.3 \\
\hline 
\rule[-1ex]{0pt}{3.5ex}  0.5 mm Cu & 8.8 $\pm$ 1.4 & 7.9 $\pm$ 1.3 & 19.0 $\pm$ 1.4 & 15.8 $\pm$ 1.3 \\
\hline
\rule[-1ex]{0pt}{3.5ex}  1.0 mm Cu & 5.6 $\pm$ 1.5 & 4.2 $\pm$ 1.4 & 13.5 $\pm$ 1.4 & 14.9 $\pm$ 1.4 \\
\hline 
\end{tabular}
\end{center}
\end{table} 

\subsection{Bin width selection}

\begin{figure}
\begin{center}
\begin{tabular}{c}
\includegraphics[width=12cm]{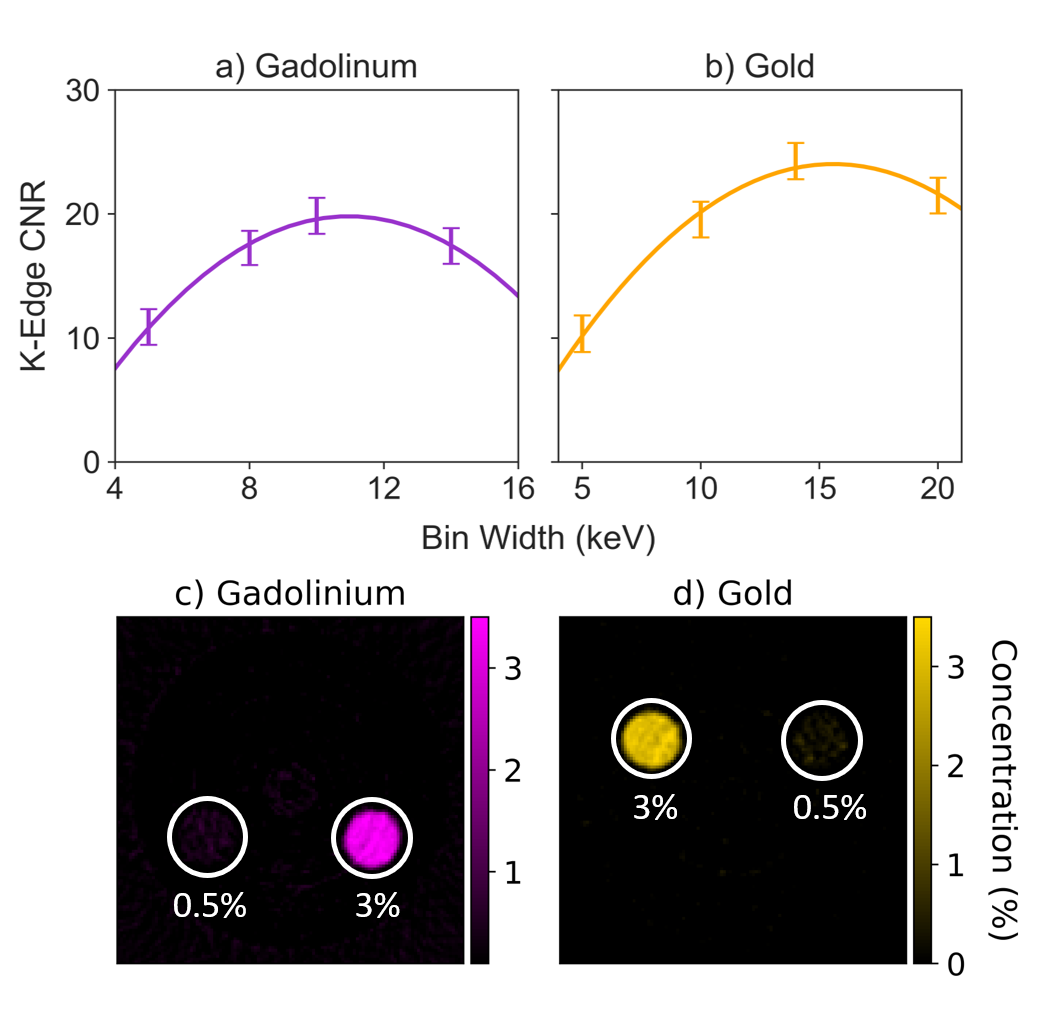}
\end{tabular}
\end{center}
\caption 
{\label{fig:Bin}
\emph{Bin selection.} a) K-edge CNR curve for gadolinium. b) K-edge CNR curve for gold. c) K-edge image of gadolinium reconstructed with 10 keV bin widths. d) K-edge image of gold reconstructed with 14 keV bin widths.} 
\end{figure}

Figure~\ref{fig:Bin}a and~\ref{fig:Bin}b demonstrate how bin width affects the CNR of gadolinium and gold K-edge images, respectively. . The data was fit with a quadratic curve to show how the K-edge CNR would likely behave over the range of bin width data that was collected. The ideal bin width must be found for every contrast material separately, as the bin width that results in the highest K-edge CNR can vary, as demonstrated. For gadolinium, peak K-edge CNR was achieved with a bin width of 10.7 keV, and for gold, the optimal bin width was 15.8 keV.  The resulting K-edge images reconstructed using the data closest to these peaks (10 keV for gadolinium and 14 keV for gold) are shown in figure~\ref{fig:Bin}c--d. The 3\% vial of each contrast can be seen clearly. However, the 0.5\% vials were not visible, with CNR values under 4. The mixed vial was not resolved in either image. 

\subsection{Projection time selection}

\begin{figure}
\begin{center}
\begin{tabular}{c}
\includegraphics[width=12cm]{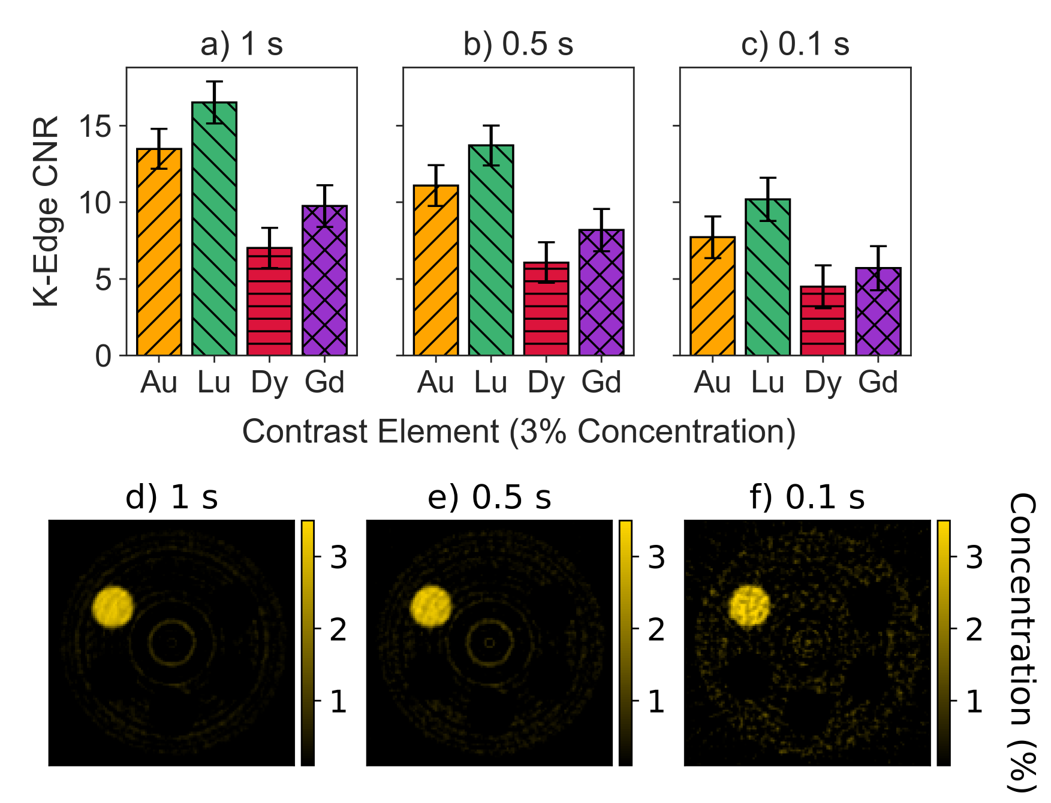}
\end{tabular}
\end{center}
\caption 
{ \label{fig:Time}
\emph{Projection time selection.)} a--c) K-edge CNR of Au, Lu, Dy, and Gd with a) 1 s acquisitions, b) 0.5 s acquisitions, c) 0.1 s acquisitions. d--f) K-edge images of gold at d) 1 s acquisitions, e) 0.5 s acquisitions, f) 0.1 s acquisitions.} 
\end{figure}

The K-edge CNR for the various contrast elements for the three projection times used in this study are shown in bar plots in figure~\ref{fig:Time}a--c. The relationship seen between CNR and imaging time is not what is expected, which is that CNR should increase proportionally to the square root of dose, or time. Figure~\ref{fig:Time}d--f shows the resultant K-edge images of gold at each of the three acquisition times. The noise increased with decreasing projection acquisition time.

\subsection{Reconstructed concentration}

\begin{figure}
\begin{center}
\begin{tabular}{c}
\includegraphics[width=12cm]{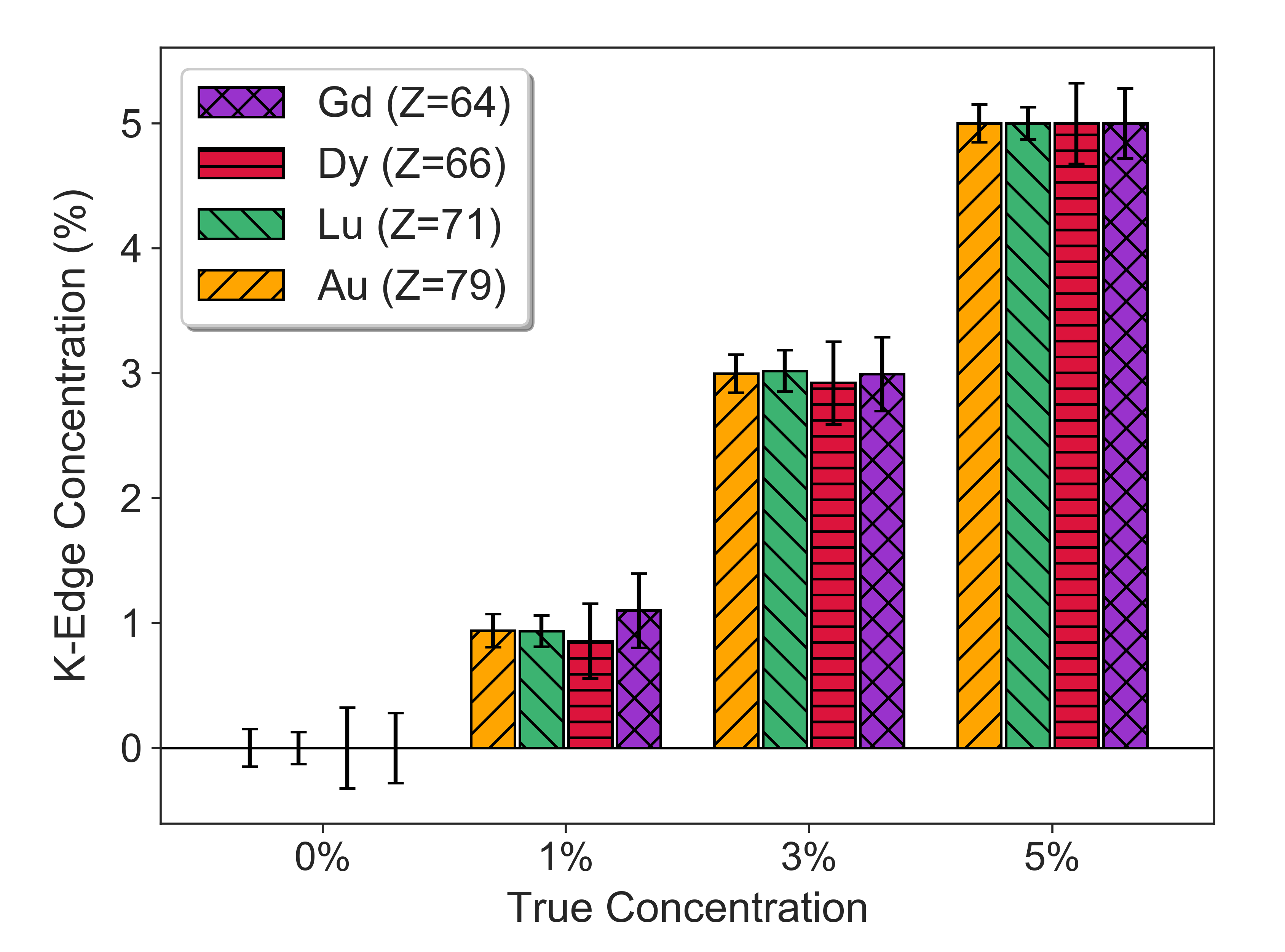}
\end{tabular}
\end{center}
\caption 
{ \label{fig:Concentration}
\emph{Reconstructed contrast concentration.} Bar graph depicting the reconstructed concentration vs. the actual concentration for four contrast agents: Au, Lu, Dy, and Gd. The values were normalized to the mean value in each of the 5\% vials.} 
\end{figure}

Figure~\ref{fig:Concentration} shows the reconstructed concentration in the K-edge images of each of the contrast elements true concentration, i.e 0, 1, 3, and 5\%. All reconstructed concentrations are accurate within the error bars when compared the true concentration. The trend of the error bar size is consistent between the contrast agents at each concentration, with a larger error for the two lower Z contrast agents (gadolinium and dysprosium) than the error for the higher Z contrast agents (lutetium and gold). The largest percent difference of the 1\% and 3\% concentrations for the four contrast agents were 0.15\%, 0.13\%, 0.32\% and 0.30\% for gold, lutetium, dysprosium, and gadolinium, respectively.

\section{Discussion}

First, we consider the contrast signal response as a function of contrast concentration seen in figure~\ref{fig:Filter}a--b. There, we see an increase in signal as Z of the contrast material increases, demonstrated by the increase in slope. This is due to the increase in attenuation coefficient that occurs as Z increases, as shown in figure~\ref{fig:Attenuation}a. In figure~\ref{fig:Attenuation}a, in the energy range from 81--120 keV, we see that the order of the attenuation coefficients follows the order of the signal in figure~\ref{fig:Filter}b. If Eq.~\ref{eq:HU} is examined, the signal (HU) at a specific energy will be higher the larger the difference is between the contrast material's linear attenuation coefficient and the linear attenuation coefficient for water. For an actual signal value, one would need to take the average attenuation of both the contrast and water, weighted by the relative number of x-rays of each energy in the relevant energy range. For the energy range from 81--120 keV, the order of the average linear attenuation coefficients is obvious, following an increase in atomic number. However, the order of attenuation coefficients in the lower energy ranges differ from what we expect, as seen in the 16--50 keV bin shown in figure~\ref{fig:Filter}a. This difference is likely caused by degradation of the energy data via a number of factors such as the energy resolution of the detector ($\sim$8 keV), charge sharing between detector elements \cite{Rajbhandary2020, Xu2011}, scattering, and fluorescent x-rays \cite{Shikhaliev2009}. Errors due to the energy resolution of the detector would affect the data by binning x-rays inaccurately because the energy of x-ray was within the energy resolution of the threshold and was binned on the opposite side of said threshold. Of these factors charge sharing likely distorted the energy data the most. Based on the analytical model by Iniewski \textit{et al.}, and extrapolating between pixel pitches of 250 and 500 $\mu$m, the expected percentage of events that suffer from charge sharing is approximately 32\% \cite{Iniewski2007}, with the lower energies being most affected. A charge sharing correction algorithm is currently under development by our collaborators. Initial results indicate, that while true counts in energy bins can be estimated accurately, resulting in reduced bias (i.e. a more accurate contrast concentration determination), noise increases and CNR decreases as a result \cite{Tanguay2020}. Ongoing work is aimed towards optimizing the bias and image noise. As a result, these preliminary corrections have not been applied here. However, with or without these corrections, it is imperative to choose the energy bin that offers the highest signal in order to best show the contrast agent when displaying CT images or to choose the bin that offers the best contrast with the other materials present.

In terms of filter selection for CT signal, shown for dysprosium and gold in figure~\ref{fig:Filter}c--d, dysprosium (figure~\ref{fig:Filter}c) showed a small filter dependence, while gold signal (figure~\ref{fig:Filter}d) had a comparatively large dependence. The order of the how the filtration affected the signal was also opposite between the two. Theoretically, based on the spectra (figure~\ref{fig:Attenuation}b), the order of the filters in the 16--50 keV bin should follow the order that gold demonstrated due to the higher number of lower energy photons that less filtration offers. This discrepancy was also likely due to effects in the lower energy range such as charge sharing.

In figure~\ref{fig:Filter K-Edge}, K-edge images for dysprosium (a--c) and gold (d--f) can be seen for each of the three filters used in this study. Image noise increased with harder filtration for dysprosium and decreased with harder filtration for gold. This noise trend can also be seen in table~\ref{tab:filterCNR}, where the K-edge CNR increased with harder filtration for contrast agents with atomic numbers below 71, while CNR decreased for contrast agents with Z-values above 71. The total number of x-ray counts was maintained over the different filters in order to keep the full spectrum image noise constant for a set imaging time. As a result, the counts in each of the different energy bins varied according to the spectra in figure~\ref{fig:Attenuation}b. Constant full spectrum image noise is desirable because the anatomical information gained from conventional CT images is necessary to localize the contrast agents. With this restriction, filter choice makes a difference on the detectability of different contrast agents based on their effective atomic number due to the relative number of counts in the two bins used to construct the K-edge images of each contrast agent. Therefore, for contrast agents with Z-values below 71, 2.0 mm Al offers the best results of the three filters tested. For those equal to or above 71, 0.5 mm Cu gives the highest K-edge CNR. One exception, based on this data, is that 0.5 mm Cu gave better CNR than 2.0 mm Al for Gd. The likely cause of this inconsistency is again the distortion in counts due to charge sharing. Since the lower energy bin used in the K-edge subtraction for Gd had a relatively higher flux with 2.0 mm Al, this could result in more charge sharing, increasing the noise in that energy bin. Additionally, the K-edge CNR with 1.0 mm Cu decreased for higher-Z contrast agents, the opposite of what should happen with higher counts. Looking at the K-edge images for gold (figure~\ref{fig:Filter K-Edge}d--f), there was a reduction in the background image noise, however the ring artifacts became more prominent, which raised the overall image noise. This was due to pixel non-uniformity that was not completely corrected by the flat field correction.

The energy bin selection shows the necessity of setting the ideal bin widths in order to maximize CNR to best separate and localize contrast agents. Meng \textit{et al.} \cite{Meng2016} stated that to get the best contrast resolution, the signal to noise ratio (SNR) must be maximized, and developed an algorithm to accomplish this. The first contributing factor to SNR in K-edge images would be the signal, which is determined by difference in the average linear attenuation coefficient of the contrast agent in the bin above the K-edge and the bin below it. The second factor affecting K-edge SNR is, of course, noise. Theoretically, this is determined by the number of x-rays falling in each of the two bins on either side of the K-edge. According to Poisson statistics, the relative noise is equal to the square root of the average number of photons over the average number of photons. So, if photon flux is constant, adjusting the bin width results in a different number of photons that would fall in that energy range. Ideally, in terms of the signal, the smallest bin size possible would be best to get the largest difference in linear attenuation and thus the most signal in the K-edge image. However, this results in a very small number of x-rays, increasing the noise significantly. Thus, a balance between the two must be found to maximize the SNR. The differences in the bin width that maximizes gadolinium CNR versus gold CNR can likely be explained by the relative difference in x-ray fluence in the bins around the respective K-edges. The x-ray spectrum filtered with 0.5 mm Cu had more x-rays around the K-edge of gadolinium (50.2 keV) than around the K-edge of gold (80.7 keV), as shown in figure~\ref{fig:Attenuation}b. This results in the need to have a wider bin for gold, relative to that for gadolinium in order to increase the counts and reduce the noise to maximize CNR. According to figure~\ref{fig:Bin}a--b, the bin widths that would result in peak CNR for gadolinium and gold would be 10.7 keV and 15.8 keV, respectively, as determined by the quadratic fit. In addition to reducing noise, the wider optimal bins are also likely a result of the $\sim$8 keV resolution of the detector. Bin widths smaller than 8 keV are likely comprised of fewer counts than what truly falls within that energy range. These optimal bin widths of 10.7 and 15.8 keV are approximately 20\% of the K-edge energy of the respective contrast agent. Using 20\% of the K-edge energy results in bin width values of 10.0 keV for gadolinium and 16.1 keV for gold. This gives a good rule-of-thumb for setting the bin widths for these contrast agents, with the 20\% values falling within 1 keV of the peak value determined from the quadratic fit.

The effect of lowering the projection acquisition time would have on K-edge CNR was also investigated for the various contrast materials. Lowering the projection acquisition time is desirable in order to keep the dose, as well as imaging time, as low as possible. Figure~\ref{fig:Time}a--c shows this effect on K-edge CNR and figure~\ref{fig:Time}d--f demonstrates how the noise increases with decreasing time. At 3\% concentration, all contrast agents were visible even at 0.1 s projection acquisitions. However, for some agents (i.e. dysprosium), no concentrations lower than 3\% would likely be resolved in K-edge images using 0.1 s acquisitions. The Rose Criterion \cite{Rose1975} states that features with SNR values of less than 5 are not readily resolved, so concentrations of dysprosium lower than 3\% would fall below a CNR of 5. Note that bin width could be chosen to raise the CNR if fewer contrast agents needed to be separated or more energy thresholds were available. In addition, theoretically, CNR should increase according to the square root of image acquisition time. This is not case in our system. From 0.1s to 1.0s, CNR should increase by a factor of 3.2, while the data only shows an increase of $\sim$1.7 times on average over all of the contrast agents. This relationship is not as expected, which is likely due to non-uniform pixel response. The relative differences in pixel responses remain constant with an increasing number of counts and thus with increasing projection time. So while increasing the projection time decreases the noise according to Poisson statistics, the near-constant non-uniform pixel response plays a major role in increasing the noise and thus depressing CNR. Unfortunately, attempting to account this using a flat field correction, as done according to equation~\ref{eq:intensity}, does not fully correct the non-uniform pixel response. This could be due in part to the relative difference in charge-sharing counts between the flat field and phantom images. For small animal studies or for other uses more research needs to be done in order to determine typical concentrations of contrast agents found in tissues in order to optimize the projection acquisition in terms of what CNRs would be measurable to keep the dose as low as reasonably possible. For future use of the bench-top PCCT system as is, the particular application will need to be considered in order to determine how long projection acquisition times should be.

Finally, the PCCT system was able to reliably determine the real concentration of the various contrast agents. Accurate determination of concentration is necessary for further use of spectral CT in small animal studies and in the clinic as it allows for the tracking of the exact amount of contrast agent or other high-Z material in various tissues. This allows for researchers investigating new therapies and imaging agents to accurately predict toxicities and the effects of their materials without \textit{ex vivo} analysis, lowering the number of animals necessary for studies and thus lowering study costs.

\section{Conclusions}
It has been demonstrated here how parameters in an experimental PCCT bench-top system can be varied to determine the configuration that offers the best imaging performance for a certain contrast agent. The PCCT system was also able to separate four different contrast agents in a range of Z-values (64--79) at varying concentrations using K-edge subtraction, even without ideal parameter selection. Possible further improvement of results could be obtained through algorithmic correction of the various mechanisms of energy data distortion, such as charge sharing. Work on charge sharing corrections for this detector are currently underway, though the results are mixed, and the algorithm was not applied here. For future use, filter type and thickness will be considered based on Z-value of the contrast agent. The 2.0 mm Al filter will be used for contrast agents with atomic numbers of less than 71, and 0.5 mm Cu used for contrast agents with Z-values of 71 and above. For energy bin width, 20\% of the contrast's K-edge energy will be used for gadolinium and gold, and verified for other contrast agents before use. Projection acquisition time will need be considered in context of the imaging task, i.e. what the limiting dose to the animal or patient is, or the desired image quality. This study demonstrates how parameters in a bench-top system can be evaluated in order to obtain the best image quality possible. 

\acknowledgments

The authors would like to thank Adriaan Frencken and Frank van Veggel for their help in synthesizing the stock lanthanide solutions. The authors would like to thank Jesse Tanguay for his assistance with charge correction algorithm development. This work was partly funded by NSERC Engage and Engage Plus grants, NSERC CGSM, an NSERC Discovery grant, the Canada Foundation for Innovation, the British Columbia Knowledge Development Fund, and the Canada Research Chair program.

\providecommand{\href}[2]{#2}\begingroup\raggedright\endgroup


\end{document}